\documentclass[preprint,prd,nofootinbib,tightenlines,amsmath]{revtex4}
\usepackage{graphicx}
\usepackage{bm}

\begin{document}
\baselineskip=15pt \parskip=5pt

\vspace*{3em}

\title{Scalar Dark Matter and Standard Model with Four Generations}

\author{Xiao-Gang He$^{1,2}$}
\author{Shu-Yu Ho$^1$}
\author{Jusak Tandean$^1$}
\author{Ho-Chin Tsai$^1$}
\affiliation{$^1$\mbox{Department of Physics, Center for Theoretical Sciences, and LeCosPA Center}, \\
National Taiwan University, Taipei \vspace*{1ex} \\
$^2$\mbox{Institute of Particle Physics and Cosmology, Department of Physics}, \\
Shanghai Jiao Tong University, Shanghai}

\date{\today $\vphantom{\bigg|_{\bigg|}^|}$}

\begin{abstract}
We consider a scalar dark matter model, the SM4+D, consisting of the standard model with four
generations (SM4) and a real gauge-singlet scalar called darkon, $D$,
as the weakly interacting massive particle (WIMP) dark-matter (DM) candidate.
We explore constraints on the darkon sector of the SM4+D from WIMP DM direct-search experiments,
including CDMS\,II and CoGeNT, and from the decay of a~$B$~meson into a~kaon plus missing energy.
We find that a~sizable portion of the darkon parameter space is still compatible with
the experimental data.
Since the darkon-Higgs interaction may give rise to considerable enhancement of the Higgs
invisible decay mode, the existence of the darkon could lead to the weakening or evasion
of some of the restrictions on the Higgs mass in the presence of fourth-generation quarks.
In addition, it can affect the flavor-changing decays of these new heavy quarks into a lighter
quark and the Higgs boson, as the Higgs may subsequently decay invisibly.
Therefore we also study these flavor-changing neutral transitions involving the darkon,
as well as the corresponding top-quark decay \,$t\to c DD$,\, some of which may be observable
at the Tevatron or LHC and thus provide additional tests for the SM4+D.
\end{abstract}

\maketitle

\section{Introduction}

The existence of dark matter (DM) in the Universe is now widely accepted.
Various observations have established that DM makes up 23\% of the total cosmic energy
density~\cite{pdg}.
Despite this evidence, however, the identity of the basic constituents of DM has
so far remained a mystery.
It is therefore important to explore different possible DM scenarios.

One of the popular candidates for DM is the weakly interacting massive particle (WIMP).
To account for WIMP DM, the standard model (SM) of particle physics needs to be enlarged.
The simplest extension of the SM possessing a WIMP candidate is the SM+D, which combines
the SM with a real SM-singlet scalar field $D$, dubbed darkon, to play the role of the DM.
This darkon model and some variations of it have been much studied in the
literature~\cite{Silveira:1985rk,McDonald:1993ex,Burgess:2000yq,darkon,Cynolter:2004cq,
Bird:2004ts,Barger:2007im,sm+d,sm+d2,Yeghiyan:2009xc,Kim:2009qc}.

In this paper we explore a somewhat enlarged darkon model we call SM4+D, which consists
of the darkon and the SM extended by the inclusion of a fourth sequential generation of
quarks and leptons.
This SM with four generations (SM4) has received much attention in recent
years~\cite{Holdom:2009rf,Maltoni:1999ta,Kribs:2007nz,Chanowitz:2009mz,Bobrowski:2009ng,
Soni:2008bc,Eilam:2009hz,Arhrib:2002md,Hou:2008xd,Frampton:1999xi,He:1985qp}.
Among the reasons~\cite{Holdom:2009rf} that have been put forward for all this interest in
the SM4 are that it is not ruled out by electroweak precision
tests~\cite{Maltoni:1999ta,Kribs:2007nz,Chanowitz:2009mz},
offers possible resolutions for certain anomalies in flavor-changing
processes~\cite{Bobrowski:2009ng,Soni:2008bc,Eilam:2009hz,Arhrib:2002md},
and might solve baryogenesis-related problems~\cite{Hou:2008xd}.
In view of the desirable features of the model, some of which nevertheless remain open 
questions, it is of interest also to consider integrating the darkon field into~it, 
assuming that the new fermions are all unstable, in which case the SM4+D is the simplest 
WIMP DM model in the presence of the fourth generation.
As we will elaborate later, the DM sector of the SM4+D can have important implications which
are absent or suppressed in the SM+D with three generations (hereafter referred to as SM3+D).
In particular, now that the LHC is operational, the extra fermions could give rise to processes
involving the darkon which are potentially observable after the LHC reaches full capacity
in the near future.

In the next section we describe the main features of the SM4+D relevant to our study.
Subsequently, after specifying the masses of the fourth-generation fermions, 
we extract the values of the darkon-Higgs coupling, to be used in later sections.
In Sec.\,\ref{direct} we explore constraints on this darkon model from
DM direct searches at underground facilities.
Recently there have been a number of such searches which can provide limits on some of
the parameter space of the darkon model.
We proceed in Sec.\,\ref{higgs} to discuss the complementarity of DM direct-detection
experiments and Higgs studies at colliders in probing the darkon properties.
The simultaneous existence of the darkon and 4th-generation fermions in the SM4+D can
have substantial impact on Higgs collider searches.
Since ongoing and near-future DM direct-search experiments are not likely to be sensitive 
to darkon masses of a few GeV or less, other processes are needed to probe the model in 
this low-mass region.
In Sec.\,\ref{b2kdd} we consider such processes, focusing on the $B$-meson decay into
a~kaon and a pair of darkons, \,$B\to KDD$,\, which contributes to
the $B$ decay into $K$ plus missing energy, \,$B\to K\mbox{$\not{\!\!E}$}$.\,
There is currently experimental information on the latter decay which can be used to
place restrictions on part of the darkon low-mass region.
In~Sec.\,\ref{q'2qdd} we explore some implications of the new fermions for the darkon sector
that are lacking or missing in the~SM3+D.
Specifically, we look at the Higgs-mediated flavor-changing top-quark decay \,$t\to c DD$,\, 
which is very suppressed in the SM3+D and can be greatly enhanced by the new-quark 
contribution, and also deal with the corresponding decays of the 4th-generation quarks.
These processes may be detectable at currently running or future colliders and, if observed,
could offer additional means to probe darkon masses from zero up to hundreds of~GeV.
We give our conclusions in Sec.\,\ref{concl}.

Before proceeding to the next section, we would like to summarize the relic-density requirements
that any WIMP candidate has to meet.
For a given interaction of the WIMP with SM4 particles, its annihilation rate into the latter and
its relic density $\Omega_D^{}$ can be calculated and are related to each other by the thermal
dynamics of the Universe within the standard big-bang cosmology~\cite{Kolb:1990vq}.
To a good approximation,
\begin{eqnarray} \label{oh}
\Omega_D^{} h^2 \,\,\simeq\,\,
\frac{1.07\times 10^9\, x_f^{}}{
\sqrt{g_*^{}}\, m_{\rm Pl}\,\langle\sigma_{\rm ann}^{}v_{\rm rel}^{}\rangle\rm\,GeV} \,\,,
\hspace{2em}
x_f^{} \,\,\simeq\,\,
\ln\frac{0.038\,m_{\rm Pl}\,m_D^{}\,\langle\sigma_{\rm ann}^{}v_{\rm rel}^{}\rangle}{
\sqrt{g_*^{}\, x_f^{}}} \,\,,
\end{eqnarray}
where  $h$ is the Hubble constant in units of 100\,km/(s$\cdot$Mpc),
\,$m_{\rm Pl}^{}=1.22\times10^{19}$\,GeV\, is the Planck mass,
\,$x_f^{}=m_D^{}/T_f^{}$\, with $T_f^{}$ being the freezing temperature, $g_*^{}$ is the number
of relativistic degrees of freedom with masses less than $T_f^{}$, and
\,$\langle\sigma_{\rm ann}^{}v_{\rm rel}^{}\rangle$\, is the thermally averaged product of
the annihilation cross-section of a pair of WIMPs into SM4 particles and the relative speed of
the WIMP pair in their center-of-mass (cm) frame.
Since $\Omega_D$ is known from observations, using the above relations one can
extract the allowed range of $\sigma_{\rm ann}^{}$ for each value of~$m_D^{}$.

\section{Brief description of SM4+D}

Being a WIMP DM candidate, the darkon $D$ has to be stable against decaying into SM4 particles.
This can be realized by assuming $D$ to be a~singlet under the SM4 gauge groups and
introducing a discrete $Z_2$ symmetry into the model.
Under the $Z_2$ transformation, \,$D\to-D$,\, while all SM4 fields are unchanged.
Requiring, in addition, that the darkon interactions be renormalizable implies that $D$ can
interact with the SM4 fields only through its coupling to the Higgs-doublet field~$H$.
It follows that the general form of the darkon Lagrangian, besides the kinetic part
\,$\frac{1}{2}\partial^\mu D\,\partial_\mu^{}D$\, and the SM4 terms, can be expressed
as~\cite{Silveira:1985rk,McDonald:1993ex,Burgess:2000yq}
\begin{eqnarray}  \label{DH}
{\cal L}_D^{} \,\,=\,\,
-\frac{\lambda_D^{}}{4}\,D^4-\frac{m_0^2}{2}\,D^2 - \lambda\, D^2\,H^\dagger H ~,
\end{eqnarray}
where  $\lambda_D^{}$,  $m_0^{}$, and $\lambda$  are free parameters, and we have followed
the notation of Ref.~\cite{sm+d2}.
The parameters in the potential should be chosen such that $D$ does not develop a vacuum
expectation value and the $Z_2$ symmetry is not broken, which will ensure that the darkon
does not mix with the Higgs field, avoiding possible fast decays into other SM4 particles.

The Lagrangian in Eq.~(\ref{DH}) can be rewritten to describe the interaction of the physical
Higgs boson $h$ with the darkon as\footnote{Obviously, $h$ here is not to be confused
with the Hubble constant, also denoted by~$h$, in the combination $\Omega_D h^2$.}
\begin{eqnarray} \label{ld}
{\cal L}_D^{} \,\,=\,\, -\frac{\lambda_D^{}}{4}\,D^4-\frac{\bigl(m_0^2+\lambda v^2\bigr)}{2}\,D^2
- \frac{\lambda}{2}\, D^2\, h^2 - \lambda v\, D^2\, h \,\,,
\end{eqnarray}
where  \,$v=246$\,GeV\,  is the vacuum expectation value of $H$.
The second term in ${\cal L}_D^{}$ contains the darkon mass
\,$m_D^{}=\bigl(m^2_0+\lambda v^2\bigr){}^{1/2}$,\, and the last term,  \,$-\lambda v D^2 h$,\,
has a major role in the determination of relic density of the darkon.
Clearly this model has a small number of free parameters in its DM sector: the darkon
mass~$m_D^{}$, the Higgs-darkon coupling~$\lambda$, and the darkon self-interaction
coupling~$\lambda_D^{}$, besides the Higgs mass~$m_h^{}$.
Our analysis will not involve~$\lambda_D^{}$.

For \,$m_D^{}<m_h^{}$\, the relic density results, at leading order, from the annihilation of
a darkon pair into SM4 particles via Higgs
exchange~\cite{Silveira:1985rk,McDonald:1993ex,Burgess:2000yq},
namely \,$DD\to h^*\to X$,\, where $X$ indicates SM4 particles.
Since the darkon is cold DM, its speed is nonrelativistic, and so a darkon pair has an invariant
mass  \,$\sqrt s\simeq2m_D^{}$.\,
With the SM4+D Lagrangian determined, the $h$-mediated annihilation cross-section of a darkon
pair into SM4 particles is then given by~\cite{Burgess:2000yq}
\begin{eqnarray} \label{csan}
\sigma_{\rm ann}^{}\, v_{\rm rel}^{} \,\,=\,\,
\frac{8\lambda^2 v^2}{\bigl(4m_D^2-m_h^2\bigr)^2+\Gamma^2_h\,m^2_h}\,
\frac{\sum_i\Gamma\bigl(\tilde h\to X_i\bigr)}{2m_D} \,\,,
\end{eqnarray}
where  \,$v_{\rm rel}^{}=2\bigl|\bm{p}_D^{\rm cm}\bigr|/m_D^{}$\, is the relative speed of
the $DD$ pair in their cm frame, $\tilde h$  is a virtual Higgs boson having
the same couplings to other states as the physical $h$ of mass $m_h^{}$, but with an invariant
mass  \,$\sqrt s=2m_D^{}$, and \,$\tilde h\to X_i$\, is any kinematically possible decay mode
of~$\tilde h$.
To determine \,$\Sigma_i\Gamma\bigl(\tilde h\to X_i\bigr)$,\, one computes
the $h$ width and then sets $m_h^{}$ equal to  $2m_D^{}$.
For \,$m_D^{}\ge m_h^{}$,\, darkon annihilation into a pair of Higgs bosons, \,$DD\to hh$,\,
also contributes to $\sigma_{\rm ann}^{}$, through $s$-, $t$-, $u$-channel, and contact diagrams
with vertices arising from the last two terms of ${\cal L}_D$ in Eq.~(\ref{ld}) and the Higgs
self-interaction (see, e.g., Ref.~\cite{Cynolter:2004cq}).
This becomes one of the leading contributions to $\sigma_{\rm ann}^{}$, along with
\,$DD\to h^*\to WW,ZZ$,\, if \,$m_D^{}\gg m_{W,Z,h}^{}$\,~\cite{McDonald:1993ex,Burgess:2000yq}.

Compared to the SM3+D case, one effect of the fourth generation of quarks and leptons in
the SM4+D is to enlarge the Higgs total width, $\Gamma_h^{}$, and also the total width
\,$\Sigma_i\Gamma\bigl(\tilde h\to X_i\bigr)$\, of the virtual Higgs,~$\tilde h$.
These new heavy fermions contribute to the total widths mainly via the decay modes into
fermion-antifermion pairs if kinematically possible and, exclusively for the new quarks,
the decay mode into a gluon pair induced by a quark loop.
Needless to say, the changes caused by the presence of these fermions depend on their masses.

There are constraints on the masses of the 4th-generation fermions from currently available
experimental data.
The masses of the heavy charged-lepton $\ell'$ and heavy neutrino $\nu'$, both assumed to be
unstable, have the PDG lower bounds \,$m_{\ell'}^{}>100.8$\,GeV\, and
\,$m_{\nu'}^{}>90.3$\,GeV\,~\cite{pdg}.
For the masses of the new up- and down-type quarks, $t'$ and $b'$, respectively, the strongest
limits are  \,$m_{t'}^{}>311$\,GeV\, and \,$m_{b'}^{}>338$\,GeV,\, from searches at
the Tevatron~\cite{Lister:2008is}.
The mass differences between the new quarks and between the new leptons turn out to be
subject to empirical constraints as well.
Electroweak precision data prefer
\,$m_{t'}^{}-m_{b'}^{}\simeq\bigl[5+\ln\bigl(m_h^{}/115{\rm\,GeV})\bigr]$$\times$10\,GeV\, and
\,$30{\rm\,GeV}\;\mbox{\small$\lesssim$}\;m_{\ell'}^{}-m_{\nu'}^{}\;
\mbox{\small$\lesssim$}\;60$\,GeV\,~\cite{Kribs:2007nz}.
Accordingly, for numerical work in this paper we take for definiteness
\,$m_{\ell'}^{}=200$\,GeV,\, \,$m_{\nu'}^{}=150$\,GeV,\, \,$m_{t'}^{}=m_{b'}^{}+55$\,GeV,
and \,$m_{t'}^{}=500$\,GeV,\, but we also sometimes make comparisons with
the \,$m_{t'}^{}=400$ and 600~GeV\, cases.
We remark that these $m_{t'}^{}$ values fall within the ranges allowed by recent global fits
for the~SM4~\cite{Kribs:2007nz,Chanowitz:2009mz,Bobrowski:2009ng,Soni:2008bc,Eilam:2009hz},
although \,$m_{t'}=600$\,GeV\, is slightly above the unitarity upper-bound
of~\,$\sim$$550$\,GeV\,~\cite{Holdom:2009rf}.

\begin{figure}[b]
\includegraphics{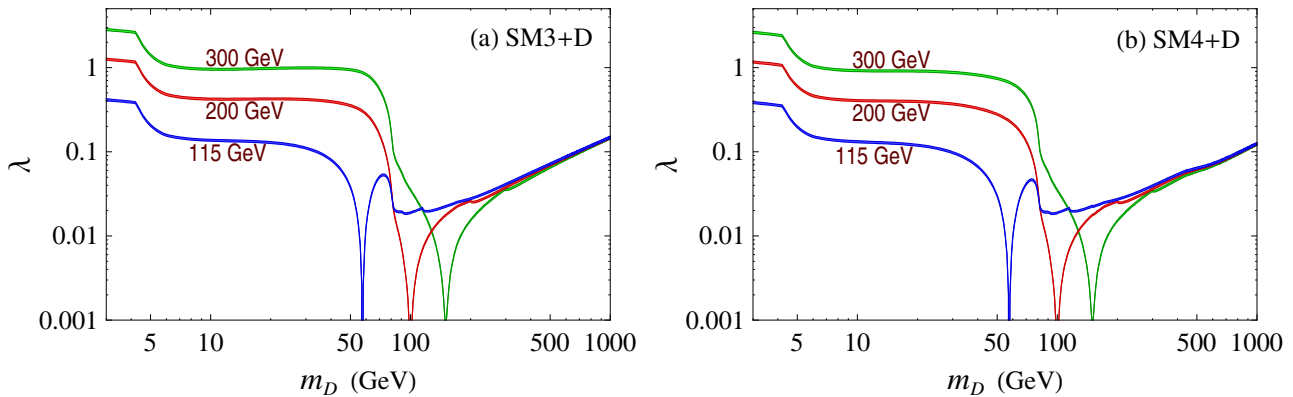}
\caption{Darkon-Higgs coupling $\lambda$ as a function of darkon mass $m_D^{}$ for Higgs mass
values \,$m_h^{}=115,200,300$\,GeV\, in (a)~SM3+D and (b)~SM4+D with \,$m_{t'}^{}=500$\,GeV.\,
The band widths in all figures correspond to the relic-density range which we have taken,
\,$0.1065\le\Omega_D^{}h^2\le0.1181$.\label{lambda-md}}
\end{figure}

With these mass choices, we can find the Higgs total widths, which we subsequently
apply in Eq.~(\ref{csan}), combined with the \,$DD\to hh$\, contribution if
\,$m_D^{}\ge m_h^{}$,\,  in order to extract the darkon-Higgs coupling $\lambda$ for given
values of $m_D^{}$, $m_h^{}$, and~\,$\langle\sigma_{\rm ann}^{}v_{\rm rel}^{}\rangle$.\,
The allowed range of \,$\langle\sigma_{\rm ann}^{}v_{\rm rel}^{}\rangle$\, as a function of
$m_D^{}$ can be inferred, with the aid of~Eq.~(\ref{oh}), from the data on the relic
density.
Its most recent value is  \,$\Omega_D^{}h^2=0.1123\pm 0.0035$,\, determined by an analysis
of the seven-year data from WMAP combined with other data~\cite{wmap7}.
From this number, one can derive the 90\%-C.L. range \,$0.1065\le\Omega_D^{}h^2\le0.1181$,\,
which we adopt for our numerical study.
We show in Fig.\,\ref{lambda-md} the resulting ranges of $\lambda$, taken to be positive,
corresponding to
\,$3{\rm\,GeV}\le m_D^{}\le1{\rm\,TeV}$\,  for some specific values of the Higgs mass,
which we choose to be \,$m_h^{}=115$, 200, and 300~GeV\, for illustration.
We present plots both in the SM3+D and in the SM4+D with \,$m_{t'}^{}=500$\,GeV\,
for comparison purposes.  The SM4+D plots for \,$m_{t'}^{}=400$ and 600~GeV\, turn out
to be very similar to the one displayed.

There are several points worth pointing out in relation to what can be seen in
Fig.\,\ref{lambda-md}.
First, although only a~relatively narrow range of the DM relic density is allowed, evidently
it can be fairly easily reproduced in both the SM3+D and~SM4+D.
Second, $\lambda$ is not small for the lower values of~$m_D^{}$, and this will result in
a considerable branching ratio of the Higgs invisible decay mode in the two models,
as we will discuss further later.
Third, for \,$3{\rm\,GeV}\mbox{\,$\le$\,}m_D^{}\mbox{\small\,$\lesssim$\,}5$\,GeV\,
the size of $\lambda$ can exceed unity and the \,$m_h^{}=300$\,GeV\, curve approaches 3
at~\,$m_D^{}\sim3$\,GeV.\,
This may seem to signal the breakdown of perturbativity in the low-$m_D^{}$ range, but
an investigation into the perturbative unitarity of darkon-Higgs interactions at tree
level~\cite{Cynolter:2004cq} has come up with the limit \,$|\lambda|<4\pi\simeq12.6$.\,
Furthermore, it has been suggested in Ref.~\cite{Bird:2004ts} that, due to a~lack of clear
division between the perturbative and nonperturbative regions of the parameter space,
a reasonable requirement is \,$|\lambda|<2\sqrt\pi\,(m_h^{}/100{\rm\,GeV})^2$,\,
which is roughly comparable to the preceding limit for the Higgs masses we have picked.
Fourth, although the $\lambda$ values tend to become small as $m_D^{}$
enters the region between 50 and 200~GeV or so, they get large again,
approximately linearly with $m_D^{}$, as $m_D^{}$ grows sufficiently large.
This follows from the facts that  \,$\langle\sigma_{\rm ann}^{}v_{\rm rel}^{}\rangle$\,
is roughly constant for the $m_D^{}$ range of interest and that
\,$\sigma_{\rm ann}^{}v_{\rm rel}^{}\simeq\lambda^2/\bigl(4\pi m_D^2\bigr)$\,
for \,$m_D^{}\gg m_{W,Z,h,t'}^{}$\,~\cite{McDonald:1993ex,Burgess:2000yq,sm+d2}.
Lastly, the curves in the SM4+D appear quite similar to the corresponding ones in the SM3+D,
but at most of the $m_D^{}$ values considered we find the former to be lower than the latter.
This decrease is mainly less than 20\%, but it reaches nearly 25\% at \,$m_D^{}\sim60$\,GeV.\,
The reason for the decrease is that the Higgs total width in the SM4 is, as mentioned earlier,
enlarged relative to that in the SM3, which is also true for the total width of $\tilde h$
in~Eq.~(\ref{csan}).
The enlargement ranges mostly from a few percent to~\,$\sim$40\%\,
and gets as high as~\,$\sim$70\%\, at~\,$m_h^{}=2m_D^{}\sim120$\,GeV.\,

\section{Constraints from dark-matter direct searches\label{direct}}

A number of underground experiments have been and are being performed to detect DM directly
by looking for the recoil energy of nuclei caused by the elastic scattering of a WIMP off
a~nucleon~\cite{Angloher:2002in,Angle:2007uj,Lin:2007ka,cdms,Aalseth:2010vx,Irastorza:2009qh}.
Although indirect DM searches have recently turned up some intriguing results which may be
interpreted as evidence for WIMPs~\cite{deBoer:2008iu}, it is very difficult to establish
a firm connection to DM due to the indirect nature of the observed events.
Therefore, direct detection is crucial to determine the properties of~DM.

In the SM4+D, the WIMP-nucleon interaction occurs via the exchange of a Higgs boson between
the darkon and the nucleon $N$ in the $t$-channel process \,$DN\to DN$.\,
Thus to evaluate this elastic scattering requires knowing not only the darkon-Higgs
coupling~$\lambda$, but also the Higgs-nucleon coupling $g_{NNh}^{}$, which parametrizes
the Higgs-nucleon interaction described by  \,${\cal L}_{NNh}^{}=-g_{NNh}^{}\,\bar NN\,h$.\,
From this Lagrangian and ${\cal L}_D^{}$ in Eq.~(\ref{ld}), one can derive for
\,$|t|\ll m^2_h$\,  the darkon-nucleon elastic
cross-section~\cite{Silveira:1985rk,McDonald:1993ex,Burgess:2000yq,Barger:2007im,sm+d}
\begin{eqnarray} \label{csel}
\sigma_{\rm el}^{} \,\,\simeq\,\,
\frac{\lambda^2\,g_{NNh}^2\,v^2\,m_N^2}{\pi\,\bigl(m_D^{}+m_N^{}\bigr)^2\, m_h^4} \,\,,
\end{eqnarray}
having used the approximation \,$\bigl(p_D^{}+p_N^{}\bigr){}^2\simeq\bigl(m_D^{}+m_N^{}\bigr){}^2$.

It remains to determine the value of $g_{NNh}^{}$, which is related to the underlying Higgs-quark
interaction described by  \,${\cal L}_{qqh}^{}=-\mbox{\large$\Sigma$}_q^{}m_q^{}\,\bar q q\,h/v$,\,
where in the SM4 the sum runs over the eight quark flavors, \,$q=u,d,s,c,b,t,b',t'$.\,
Since the energy transferred in the darkon-nucleon scattering is very small, of order tens of~keV,
one can employ a chiral-Lagrangian approach to estimate~$g_{NNh}^{}$.
This has been done previously in the context of the~SM3~\cite{sm+d,Shifman:1978zn,Cheng:1988cz}.
In the~SM4 case, we modify the derivation described in Ref.~\cite{sm+d},
incorporating the $t'$ and~$b'$ contributions, to arrive at
\begin{eqnarray}
g_{NN\cal H}^{\rm SM4} \,\,=\,\, \frac{m_N^{}\,-\,\frac{17}{27}\,m_B^{}}{v} ~,
\end{eqnarray}
where $m_N^{}$ is the nucleon mass and $m_B^{}$ denotes the lightest octet-baryon mass in
the chiral limit, which can be related to the pion-nucleon sigma term, $\sigma_{\pi N}^{}$,
by  \,$m_B^{}\simeq -13.39\,\sigma_{\pi N}^{} + 1.269$\,GeV\,~\cite{sm+d}.
With \,$\sigma_{\pi N}^{}=45$\,MeV\,~\cite{Gasser:1990ce}, we obtain
\begin{eqnarray}
g_{NNh}^{\rm SM4} \,\,=\,\, 2.11\times10^{-3} ~,
\end{eqnarray}
to be compared with the SM3 value \,$g_{NNh}^{\rm SM3}=1.71\times10^{-3}$\,~\cite{sm+d}.
We adopt these numbers in our numerical calculations below.
We note, however, that $\sigma_{\pi N}^{}$ is not well determined, with values ranging
roughly from 35\,MeV to 80\,MeV having been quoted in
the literature~\cite{Cheng:1988cz,Gasser:1990ce,Ellis:2008hf}, which translate into
\,$1.8\times10^{-3}\;\mbox{\small$\lesssim$}\;g_{NNh}^{\rm SM4}\;
\mbox{\small$\lesssim$}\;3.3\times10^{-3}$\, and
\,$1.3\times10^{-3}\;\mbox{\small$\lesssim$}\;g_{NNh}^{\rm SM3}\;
\mbox{\small$\lesssim$}\;3.2\times10^{-3}$.\,

\begin{figure}[b]
\includegraphics[width=4.4in]{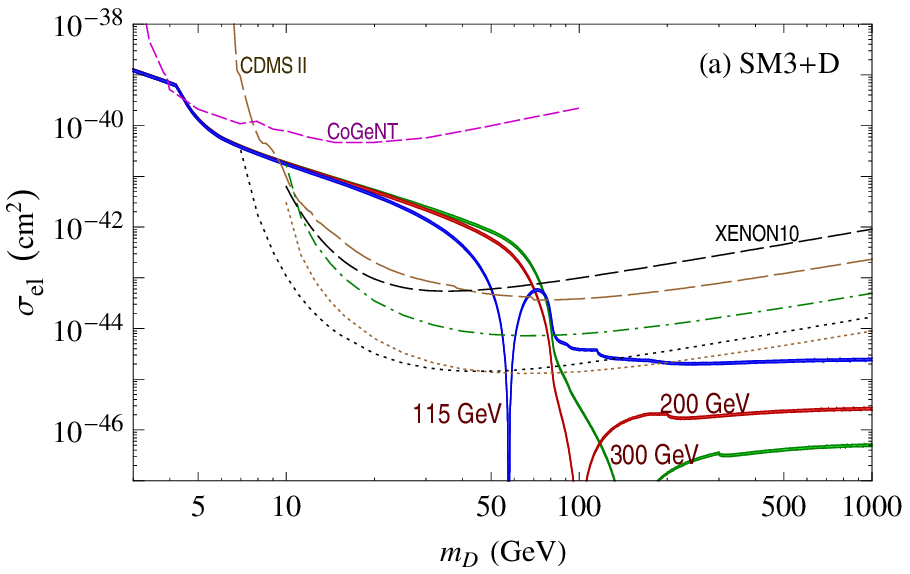} \\
\includegraphics[width=4.4in]{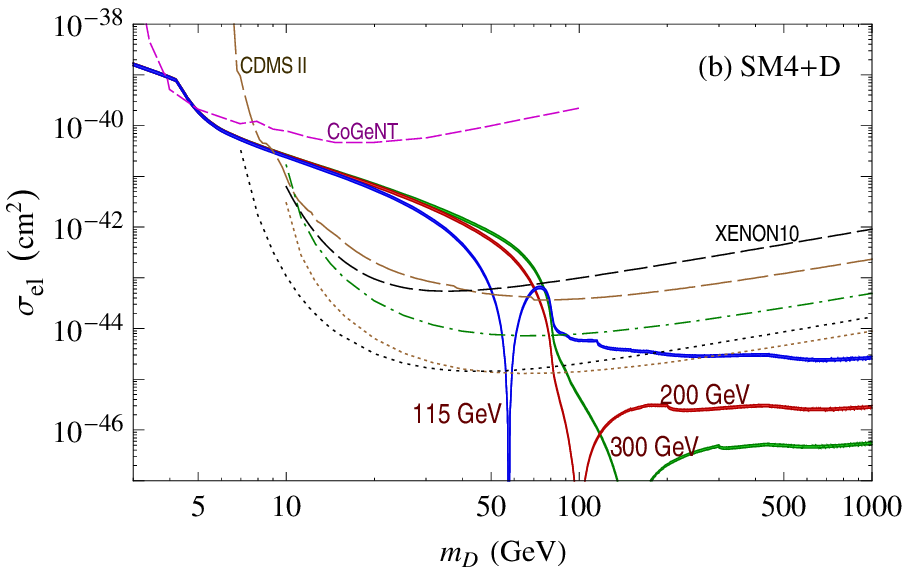} \vspace*{-1ex}
\caption{Darkon-nucleon elastic cross-section $\sigma_{\rm el}^{}$ as a function of darkon
mass $m_D^{}$ for Higgs mass values \,$m_h^{}=115,200,300$\,GeV\, in (a)~SM3+D and
(b)~SM4+D with \,$m_{t'}^{}=500$\,GeV,\, compared to 90\%-C.L. upper limits from
XENON10 (black dashed-curve), CDMS\,II (brown [gray] dashed-curve),
and CoGeNT (purple dashed-curve), as well as projected sensitivities of SuperCDMS at Soudan
(green dot-dashed curve), SuperCDMS at Snolab (brown [gray] dotted curve), and
XENON100 (black dotted curve).\label{elastic_cs}}
\end{figure}

With $\lambda$ and $g_{NNh}^{}$ known, we can now predict the darkon-nucleon elastic
cross-section $\sigma_{\rm el}^{}$ as a~function of darkon mass once the Higgs mass is specified.
We show our results for $\sigma_{\rm el}^{}$ in Fig.\,\ref{elastic_cs}, where the choices of
darkon and Higgs masses are the same as those in Fig.\,\ref{lambda-md}.
For comparison, we display $\sigma_{\rm el}^{}$ graphs in the SM3+D and in the SM4+D with
\,$m_{t'}^{}=500$\,GeV.\,
The SM4+D plots for \,$m_{t'}^{}=400$ and 600~GeV\, are again similar to the one displayed.
We find that at most of the $m_D^{}$ values considered the $\sigma_{\rm el}^{}$ curves in
the SM4+D for a fixed $m_h^{}$ are higher than the corresponding ones in the SM3+D.
This difference ranges from a~few to roughly 50~percent and results from
the $g_{NNh}^{\rm SM4}$ enhancement relative to $g_{NNh}^{\rm SM3}$  overcoming
the $\lambda_{\rm SM4+D}$ suppression relative to $\lambda_{\rm SM3+D}$ mentioned in
the preceding section.

In Fig.\,\ref{elastic_cs} we also plot the 90\%-C.L. upper-limit curves for the WIMP-nucleon
spin-independent elastic cross-section reported by the XENON10~\cite{Angle:2007uj},
CDMS\,II~\cite{cdms}, and CoGeNT~\cite{Aalseth:2010vx} experiments,
along with the expected sensitivities of a number of future experiments~\cite{dmplot}.
For \,$m_D^{}\mbox{\small\,$\lesssim$\,}10$\,GeV,\, there are also limits from
CRESST-I~\cite{Angloher:2002in} and TEXONO~\cite{Lin:2007ka}, but they are both above
the predictions of the two models.

Comparing the prediction curves of both models to the experimental upper-bounds in
Fig.\,\ref{elastic_cs}, one can see that some portions of the darkon mass regions
considered are excluded, but the greater part of them are still viable.
More precisely, for \,$m_h^{}=115,\,200$, and 300~GeV\, the XENON10 and CDMS\,II limits
have ruled out darkon masses from $\sim$9~GeV to between 70 and 80~GeV, except for
the \,$50{\rm\,GeV}\mbox{\,$<$\,}m_D^{}\mbox{\,$<$\,}70$\,GeV\, range in
the \,$m_h^{}=115$\,GeV\, case.
Moreover, in the low-$m_D^{}$ sections of the plots the exclusion limit from CoGeNT
can be seen to rule out part of the
\,$4{\rm\,GeV}\mbox{\small\,$\lesssim$\,}m_D^{}\mbox{\small\,$\lesssim$\,}5$\,GeV\, range.
In contrast, darkon masses larger than 80\,GeV or~so are not yet probed by the current data
from direct searches.
As the projected sensitivities of future experiments in this figure suggest, SuperCDMS at Snolab
and XENON100 may probe these models further to \,$m_D^{}\sim400$\,GeV,\, but SuperCDMS at Soudan
may be unlikely to provide much stronger constraints on the models than the present bounds.

It is interesting to point out that these two darkon models can accommodate the possibility that
the excess events observed by CoGeNT originate from interactions with a relatively light WIMP
of mass between 7 and 11~GeV~\cite{Aalseth:2010vx},
which is compatible with the two signal-like events detected by CDMS\,II~\cite{cdms}
if they are also interpreted as evidence for WIMP interactions.
For $m_D^{}$ values within this range, the prediction curves in Fig.\,\ref{elastic_cs}
each have some overlap with the possible signal region reported by CoGeNT~\cite{Aalseth:2010vx}.
This is more so if we take into account the uncertainties in $g_{NNh}^{}$ noted above,
which could imply an increase in the predicted $\sigma_{\rm el}^{}$ by up to a~factor of~3.

Before moving on, it is worth remarking that, as Fig.\,\ref{elastic_cs} indicates,
$\sigma_{\rm el}^{}$ for a fixed $m_h^{}$ approaches a~constant value as $m_D^{}$ becomes
much greater than $m_{W,Z,h,t'}^{}$.
The reason is that in this large-$m_D^{}$ limit the ratio \,$\lambda^2/m_D^2$\, is,
as pointed out in the preceding section, approximately constant and $\sigma_{\rm el}^{}$
in Eq.~(\ref{csel}) is proportional to the same ratio,~$\lambda^2/m_D^2$.
Another observation from this figure is that the asymptotic value of $\sigma_{\rm el}^{}$
decreases as $m_h^{}$ increases, which is in accord with Eq.~(\ref{csel}).
Hence direct DM searches in the future may lack the sensitivity to probe the larger darkon
masses if the Higgs mass is also large.

\section{Some implications for Higgs searches at colliders\label{higgs}}

Since both the SM3+D and SM4+D have only a small number of free parameters, the relevant ones
here being $\lambda$, $m_D^{}$, and $m_h^{}$,
it is possible to draw strong correlations among them~\cite{sm+d2}.
This implies that these darkon models have a high degree of predictivity and that there is
some simplification in testing them, without requiring many different observables.
To illustrate this, we now discuss the Higgs decay into a pair of darkons, \,$h\to DD$,\,
and some of its consequences for Higgs studies at colliders, in light of the bounds obtained
above from comparing with DM direct-detection data.

\begin{figure}[b]
\includegraphics{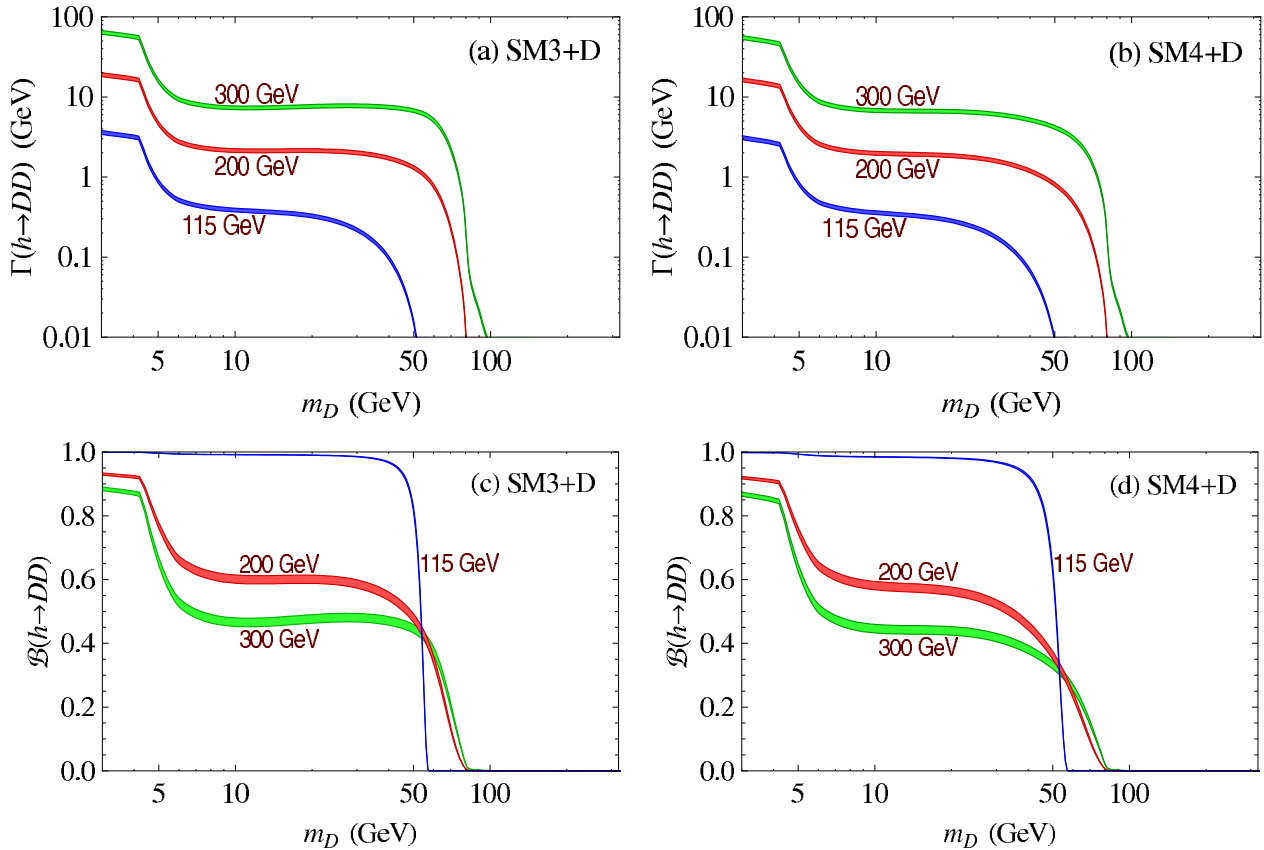}
\caption{Partial width and branching ratio of invisible decay \,$h\to DD$\, as functions
of darkon mass $m_D^{}$ for Higgs mass values  \,$m_h^{}=115,200,300$\,GeV\,
in (a,c)~SM3+D and (b,d)~SM4+D with \,$m_{t'}^{}=500$\,GeV.\label{br-md}\vspace*{-2ex}}
\end{figure}

Using the $\lambda$ values obtained in Sec.\,\ref{direct}, we compute the rate and branching
ratio of the invisible mode \,$h\to DD$.\,
The results are depicted in Fig.\,\ref{br-md}, where the Higgs and darkon mass choices are
the same as those in Fig.\,\ref{lambda-md}.
One can observe that the situations in the SM3+D and SM4+D are similar, namely that
the sizable values of $\lambda$ in Fig.\,\ref{lambda-md} translate into huge enhancement
of the Higgs width via the additional process \,$h\to DD$\, and, consequently, an invisible
branching ratio that is large.
This is especially so if \,$2m_D^{}<m_h^{}<2m_W^{}$,\, in which case the Higgs partial
width into standard particles is small.
Although the 4th-generation quarks can cause the decay mode into a gluon pair, \,$h\to gg$,\,
to dominate if \,$m_h^{}\;\mbox{\small$\lesssim$}\;140$\,GeV\,~\cite{Kribs:2007nz},
the inclusion of the darkon in the SM4+D leads to the dominance of \,$h\to DD$\, instead.

The potential importance of the Higgs invisible decay mode in the darkon presence can
alleviate some of the restrictions on Higgs masses in the SM4.
For instance, at hadron colliders the important channel \,$gg\to h\to WW\to\ell\nu\ell\nu$\,
is expected to be enhanced due to the new quarks~in the SM4 by a factor of 9
for \,$100{\rm\,GeV}\mbox{\small\,$\lesssim$\,}m_h^{}\mbox{\small\,$\lesssim$\,}200$\,GeV,\,
the measurement of which would also provide indirect evidence for the new quarks~\cite{Holdom:2009rf}.
Preliminary searches at the Tevatron for this channel have so far come back negative, thus excluding
a large portion of this $m_h^{}$ range in the SM4~\cite{Holdom:2009rf,tevatron}.
In the SM4+D, however, the possible dominance of the \,$h\to DD$\, mode implies that
the enhancement of \,$gg\to h\to WW\to\ell\nu\ell\nu$\, would likely be reduced or even negated
completely, and so the Higgs-mass constraints could be weakened or evaded.

As discussed previously~\cite{sm+d2}, collider measurements of the Higgs invisible decay
with sufficient precision can lift some possible ambiguities in determining the darkon mass
from direct DM searches.
A substantial Higgs invisible decay can also be advantageous for testing the darkon models
if the constraints from direct DM searches are combined with Higgs studies at colliders.
As found above, the greater part of the darkon mass range from about 9 to 80~GeV
in the SM3+D and SM4+D have been ruled out by direct-detection data if the Higgs mass
\,$m_h^{}=115$, $200$, or 300~GeV,\, the main exception being the neighborhood
of~\,$m_D^{}\sim57.5$\,GeV.\,
Since Fig.\,\ref{br-md} shows that a Higgs boson with one of these masses decays dominantly or
significantly into a darkon pair if \,$m_D^{}\mbox{\small\,$\lesssim$\,}50$\,GeV,\,
then the observation of such a Higgs boson with a sizable invisible branching-ratio might
hint at inconsistencies of the models.
All this illustrates that the interplay between direct DM searches and the study of
the Higgs boson at colliders can yield crucial information about the darkon properties.

An enhanced ${\cal B}(h\to DD)$ affects not only collider searches for the Higgs boson,
but also decays which are mediated by or produce it.
In the following two sections, we deal with such processes arising from the Higgs
flavor-changing neutral couplings to quarks.

\section{Constraints from \boldmath$B\to KDD$ decays\label{b2kdd}}

As seen in Sec.\,\ref{direct}, direct DM searches with underground detectors currently
being done or to be done in the near future are not expected to be sensitive to darkon masses
below a few GeV.
It turns out that such darkon masses can be probed using the decays of mesons containing
the~$b$~quark.
In this section we explore constraints available from the $B$-meson decay \,$B\to KDD$,\,
which contributes to the $B$ decay into $K$ plus missing energy,
\,$B\to K\mbox{$\not{\!\!E}$}$.\,
One could carry out a~similar analysis using \,$B\to K^*DD$,\, but we will not do so here.
We will also briefly comment on the spin-one bottomonium decay \,$\Upsilon\to DD\gamma$.\,

Since the Higgs boson $h$ is the only SM particle to which $D$ couples, \,$B\to KDD$\,
is induced by the flavor-changing $b$-quark decay \,$b\to s h^*\to sDD$,\,  the effective
$bsh$ coupling being loop-generated with up-type quarks and the $W$ boson in the loops.
These transitions have been studied previously in the context of the SM3+D in
Refs.~\cite{Bird:2004ts,Kim:2009qc}.
Generalizing their results to the SM4+D, we can express the effective Hamiltonian for
\,$b\to s h^*\to sDD$\, as
\begin{eqnarray}  \label{b2sDD}
{\cal H}_{b\to sDD}^{} \,\,=\,\,
\frac{\lambda\,g_{bs}^{}\,m_b^{}}{2 m_h^2}\, \bar s\bigl(1+\gamma_5^{}\bigr)b\,D^2 ~,
\end{eqnarray}
where
\begin{eqnarray} \label{gbs}
g_{bs}^{} \,\,=\,\, \frac{3 g^2}{64\pi^2}
\bigl(\lambda_t^{bs}\,x_t^{}+\lambda_{t'}^{bs}\,x_{t'}^{}\bigr) ~, \hspace{5ex}
x_q^{} \;=\; \frac{m_q^2}{m_W^2} ~, \hspace{5ex} \lambda_q^{bs} \;=\; V^*_{qs}V_{qb}^{} ~,
\end{eqnarray}
with $V_{kl}$ being the elements of the 4$\times$4 Cabibbo-Kobayashi-Maskawa (CKM4)
matrix, and contributions from $u$ and $c$ quarks have been neglected.
Hence the corresponding expression for $g_{bs}^{}$ in the SM3+D does not contain
the \,$\lambda_{t'}^{bs}x_{t'}^{}$\, term.

The amplitude for \,$B^-\to K^- DD$\, is then
\begin{eqnarray}  \label{M_B2KDD}
{\cal M}(B^-\to K^-DD) \,\,=\,\, \frac{\lambda\,g_{bs}^{}\,m_B^2}{m_h^2}\, f_0^{}(\hat s) ~,
\end{eqnarray}
where
\,$f_0^{}(\hat s)=0.3\,\exp\bigl(0.63\,\hat s/m_B^2-0.095\,\hat s^2/m_B^4+0.591\,\hat s^3/m_B^6\bigr)$\,
is the relevant \,$B\to K$\, form-factor~\cite{Bird:2004ts}, with
\,$\hat s=\bigl(p_B^{}-p_K^{}\bigr){}^2$,\, and the approximation
\,$\bigl(m_B^2-m_K^2\bigr)/\bigl(m_b^{}-m_s^{}\bigr)\simeq m_B^2/m_b^{}$\, has been made.
It follows that
\begin{eqnarray}  \label{wB2KDD'}
\Gamma(B^+\to K^+DD) \;=\;  \frac{\lambda^2}{m_h^4}\,
\frac{|g_{bs}^{}|^2\,m_B^{}}{512\pi^3}\,I\bigl(m_D^{}\bigr) ~,
\end{eqnarray}
where a factor of 1/2 has been included to account for the identical $D$'s in the final state and
\begin{eqnarray}
I\bigl(m_D^{}\bigr) \;=\;
\int_{4m_D^2}^{(m_B^{}-m_K^{})^2} d\hat s\;\bigl(f_0^{}(\hat s)\bigr)^2\,
\sqrt{1-\frac{4m_D^2}{\hat s}}\, \sqrt{\bigl(m_B^2-m_K^2-\hat s\bigr)^2-4m_K^2\,\hat s} ~.
\end{eqnarray}
We note that our formula for the $B\to KDD$ rate agrees with the corresponding one
obtained in Ref.~\cite{Kim:2009qc}, but is 4 times smaller than that given in
Ref.~\cite{Bird:2004ts}.\footnote{This can be traced to a factor of 1/2 apparently missing in
the expression for the \,$B\to KDD$\, amplitude in the Eq.~(6) of the first paper in
Ref.~\cite{Bird:2004ts}.}
Since  for \,$m_D^{}\ll m_h^{}$\,  we can simplify Eq.~(\ref{csan})  to
\begin{eqnarray} \label{csan'}
\sigma_{\rm ann}^{}\, v_{\rm rel}^{} \,\,\simeq\,\,
\frac{\lambda^2}{m_h^4}\, \frac{4v^2\sum_i\Gamma\bigl(\tilde h\to X_i\bigr)}{m_D} ~,
\end{eqnarray}
we can rewrite Eq.~(\ref{wB2KDD'}) as
\begin{eqnarray}  \label{wB2KDD}
\Gamma(B^+\to K^+DD) \;\simeq\;  \frac{|g_{bs}^{}|^2\,m_B^{}\,I\bigl(m_D^{}\bigr)}{2048\pi^3\,v^2}\,
\frac{\bigl(\sigma_{\rm ann}^{}v_{\rm rel}^{}\bigr)m_D^{}}
{\sum_i\Gamma\bigl(\tilde h\to X_i\bigr)} \,\,,
\end{eqnarray}
where both $\sigma_{\rm ann}^{}v_{\rm rel}^{}$ and $\Sigma_i\Gamma\bigl(\tilde h\to X_i\bigr)$
have $m_D^{}$ dependence.
We can get constraints on the parameter space \,$m_D^{}\le(m_B^{}-m_K^{})/2$\, by comparing
this prediction with the experimental information on the $B$ decay into a kaon plus missing energy,
which receives contributions from \,$B\to KDD$\, and \,$B\to K\nu\bar\nu$.\,
We first update the constraints in the SM3+D and then discuss the SM4+D case.

\begin{figure}[t]
\includegraphics[width=3.33in]{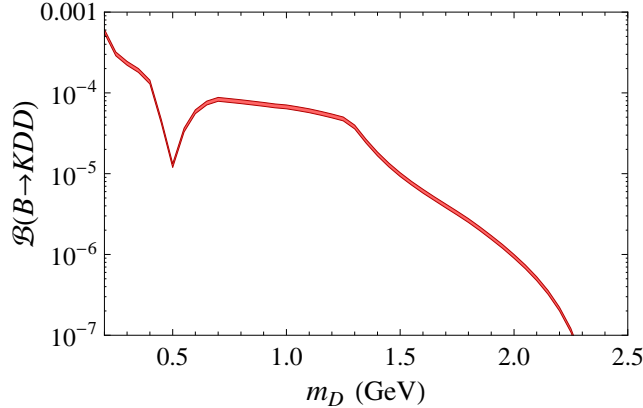} \vspace*{-1ex}
\caption{Branching ratio of \,$B^+\to K^+DD$\, as a function of darkon mass $m_D^{}$
in~SM3+D.\label{brB2KDD}}
\end{figure}

The prediction of the branching ratio ${\cal B}(B\to K DD)$ in the SM3+D using Eq.~(\ref{wB2KDD}),
with the \,$\lambda_{t'}^{bs}x_{t'}^{}$\, term in $g_{bs}^{}$ dropped,
involves large uncertainties which come mainly from the calculation of the total width
$\Sigma_i\Gamma\bigl(\tilde h\to X_i\bigr)$ for $m_{\tilde h}$ under a~few~GeV.
In the case of the physical $h$, for \,$m_h^{}\mbox{\small\,$\lesssim$\,}2$\,GeV\,
the predicted rate of the important channel \,$h\to{\rm hadrons}$\, is well known to contain
significant uncertainties~\cite{Raby:1988qf,Donoghue:1990xh}.
In estimating the $\tilde h$ total width, for \,$2m_\pi^{}\le m_{\tilde h}\le1.4$\,GeV\, we adopt
the \,$\Gamma(h\to{\rm hadrons})$\, results from Ref.~\cite{Donoghue:1990xh},
whereas for smaller and larger values of \,$m_{\tilde h}$\, we simply use the perturbative
formulas for Higgs decays~\cite{Djouadi:2005gi}.
We graph the resulting ${\cal B}(B\to K DD)$ as a~function of $m_D^{}$ in Fig.~\ref{brB2KDD},
which is to be compared with experimental data.

The latest experimental search for the decay \,$B^+\to K^+\nu\bar\nu$\, has produced
the branching-ratio limit
\,${\cal B}_{\rm exp}(B^+\to K^+\nu\bar\nu)<14\times10^{-6}$\,~\cite{:2007zk}.
On the theoretical side, the most recent calculations in the SM3 predict
${\cal B}(B^+\to K^+\nu\bar\nu)$ to be between $3.6\times10^{-6}$ and $5.1\times10^{-6}$,
with errors of order~15\%~\cite{Altmannshofer:2009ma}.
Accordingly, it is reasonable to take
\,${\cal B}_{\rm exp}(B^+\to K^+\mbox{$\not{\!\!E}$})\simeq
{\cal B}_{\rm exp}(B^+\to K^+\nu\bar\nu)$,\,
from which we can subtract the SM3 prediction for ${\cal B}(B^+\to K^+\nu\bar\nu)$ in order to
require that \,${\cal B}(B^+\to K^+DD)<1\times10^{-5}$\, in the SM3+D.

We can see from Fig.\,\ref{brB2KDD} that the predicted
\,${\cal B}(B^+\to K^+DD)>1\times10^{-5}$\, for \,$m_D^{}\;\mbox{\small$\lesssim$}\;1.5$\,GeV.\,
Recalling the hadronic uncertainties mentioned above, we can then conclude that in
the SM3+D much of this range of $m_D^{}$ values,
especially \,$m_D^{}\;\mbox{\small$\lesssim$}\;0.4$\,GeV,\, is excluded by the data.
We would need improved data from future measurements of \,$B\to K\mbox{$\not{\!\!E}$}$\,
before we could disallow more darkon masses within the \,$m_D^{}<2.4$\,GeV\, region.
These conclusions are similar to those made in Ref.~\cite{Bird:2004ts} due partly to
the stronger experimental limit at present and partly to the overestimate of
their~${\cal B}(B\to K DD)$.

In the SM4+D, the prediction for $\Gamma(B\to K DD)$ is modified due to the presence
of the new quarks, $t'$ and $b'$.
The loop-induced effective coupling $g_{bs}^{}$ in Eq.~(\ref{wB2KDD}) receives
a $t'$-quark contribution as given in Eq.~(\ref{gbs}).
To examine its effect on $\Gamma(B\to K DD)$, we need to compare $g_{bs}^{\rm SM4}$
to~$g_{bs}^{\rm SM3}$.
For concreteness, we take the relevant CKM4 elements extracted in Ref.~\cite{Eilam:2009hz}
from a~global fit for the SM4.
Accordingly, we can expect that the numbers we use are typical values for the model.
Thus, with \,$\lambda_t^{bs}=0.04$\, in $g_{bs}^{\rm SM3}$, we find
\,$|g_{bs}^{\rm SM4}/g_{bs}^{\rm SM3}|^2$\, for \,$m_{t'}^{}=400$ and 500~GeV to be similar
in value, \,$\sim$1.2,\, but it goes up to 1.6 for \,$m_{t'}^{}=600$\,GeV.\,
In addition, the presence of $t'$ and $b'$ affects the total width
$\Sigma_i\Gamma\bigl(\tilde h\to X_i\bigr)$ in Eq.~(\ref{wB2KDD}) mainly via their
contributions to \,$\tilde h\to{\rm hadrons}$\, due to the quark-loop induced
\,$\tilde h\to gg$,\, as already mentioned earlier.
Despite the hadronic uncertainties, this implies that the enhancement of the $\tilde h$ total
width in the SM4 compared to the SM3 can be expected to be less than \,$(5/3)^2\simeq2.8$\,
if \,$2m_\pi^{}\le m_{\tilde h}\le2 m_c^{}$,\, where 5 and 3 are the numbers of heavy quarks in
the two models, respectively, for this $m_{\tilde h}$ range.
This enhancement decreases to no more than 25\% after the \,$\tilde h\to c\bar c$\,
channel is open.
We can then conclude that the effects of these two factors on $\Gamma(B\to K DD)$ in the SM4+D
amount to changes to the rate in the SM3+D by no more than a factor of 2 in either direction,
implying that the curve in Fig.~\ref{brB2KDD} would not be very different in the SM4+D.
Since the prediction for ${\cal B}(B^+\to K^+\nu\bar\nu)$ is raised by at
most~20\% in the SM4~\cite{Soni:2008bc}, the empirical bound on ${\cal B}(B\to K DD)$ in
the SM4+D can be taken to be unchanged compared to that in the SM3+D.
It follows that the constraints on the darkon masses within the \,$m_D^{}<2.4$\,GeV\, range
in the SM4+D are similar to those in the SM3+D.

We should also mention that for \,$m_D^{}\mbox{\small\,$\lesssim$\,}170$\,MeV\,
an additional restriction is provided by the kaon decay \,$K^+\to\pi^+\mbox{$\not{\!\!E}$}$,\,
which receives a contribution from \,$K^+\to\pi^+DD$.\,
The agreement between the SM3 expectation and experimental data on
\,$K^+\to\pi^+\mbox{$\not{\!\!E}$}$\, implies that for
\,$m_D^{}\mbox{\small\,$\lesssim$\,}170$\,MeV\, the SM3+D is already ruled
out~\cite{Bird:2004ts}, as is the~SM4+D.
This is consistent with what can be inferred from the low-$m_D^{}$ end of Fig.~\ref{brB2KDD}.

For the larger range \,$m_D^{}\mbox{\small\,$\lesssim$\,}5$\,GeV,\, there may also be
constraints available from future measurements of the decays \,$\Upsilon\to DD\gamma$.\,
Presently, the existing experimental limits on \,$\Upsilon\to\gamma+\rm invisible$\,
are not yet strong enough to probe these darkon models~\cite{Yeghiyan:2009xc}.

\section{FCNC decays \boldmath\,$Q\to qDD$\label{q'2qdd}}

The presence of the new quarks in the SM4+D can have important implications for probing
the darkon sector that are lacking or absent in the SM3+D.
In the SM3 the flavor-changing neutral current (FCNC) top-quark decay \,$t\to c h$\, is
known to be very suppressed, with a~branching ratio estimated to be between $10^{-15}$ and
$10^{-13}$~\cite{Eilam:1990zc,AguilarSaavedra:2000aj}, but in the SM4 the branching ratio can
be enhanced by several orders of magnitude~\cite{Eilam:2009hz,Arhrib:2006pm}.
We expect that in the SM4+D the related decay  \,$t\to c h^*\to cDD$,\,
if kinematically allowed, can be similarly enhanced.
These processes may be detectable at the LHC after its operation reaches full capacity
in the near future.
The Tevatron and the LHC can also produce the new quarks, $t'$ and $b'$, if they exist,
in a similar way as they can produce the $t$ quark, albeit fewer of them due to their bigger masses.
It is therefore of interest as well to explore their FCNC decays, \,$t'\to(c,t)h^*\to(c,t)DD$\,
and \,$b'\to(s,b)h^*\to(s,b)DD$,\, which may have observable rates.
These decays could, in principle, probe darkon masses from zero all the way up to
$(m_Q^{}-m_q^{})/2$, hence covering potentially wider $m_D^{}$ ranges than those covered by
some of the DM direct searches in the future.
Here we estimate the branching ratios of these FCNC decays involving the darkon.
The corresponding decays with the $u$ and $d$ quarks,
\,$t^{(\prime)}\to uDD$\, and \,$b'\to dDD$,\, are comparatively suppressed due to the less
favorable CKM4 factors.

The Lagrangian describing the FCNC transition \,$Q\to qh$\, involving a heavy quark $Q$ and
a~lighter quark $q$ can be written as
\begin{eqnarray}  \label{u2uh}
{\cal L}_{Qqh}^{} \,\,=\,\, \bar q\bigl(g^{Qq}_{\rm L}P_{\rm L}^{} \,+\,
g^{Qq}_{\rm R}P_{\rm R}^{}\bigr)Q\,h ~,
\end{eqnarray}
where \,$P_{\rm L,R}^{}=\frac{1}{2}(1\mp\gamma_5^{})$\, and the loop-induced couplings
$g^{Qq}_{\rm L,R}$ generally depend not only on the internal quark (and $W$) masses and
the CKM matrix elements, but also on the masses and momenta of the external particles.
The amplitude for \,$Q\to qh^*\to q DD$\, is then
\begin{eqnarray}
{\cal M}(Q\to q DD) \,\,=\,\,
\frac{2\lambda v\,\bar q\bigl(g^{Qq}_{\rm L}P_{\rm L}^{}+g^{Qq}_{\rm R}P_{\rm R}^{}\bigr)Q}
     {m_h^2-\bar s-i\Gamma_h^{}m_h^{}} ~,
\end{eqnarray}
where \,$\bar s=(p_Q^{}-p_q^{})^2$.\,
This yields the decay rate
\begin{eqnarray}
\Gamma(Q\to qDD) &=&
\frac{\lambda^2\,v^2}{256\pi^3\,m_Q^3}\int_{4m_D^2}^{(m_Q^{}-m_q^{})^2} d\bar s\;
\sqrt{1-\frac{4m_D^2}{\bar s}}\,\sqrt{\bigl(m_Q^2-m_q^2-\bar s\bigr)^2-4m_q^2\bar s}
\nonumber \\ && \hspace*{11ex} \times\,\,
\frac{\Bigl(\bigl|g^{Qq}_{\rm L}\bigr|^2+\bigl|g^{Qq}_{\rm R}\bigr|^2\Bigr)\bigl(m_Q^2+m_q^2-\bar s\bigr)
      + 4\,{\rm Re}\bigl(g^{Qq*}_{\rm L}g^{Qq}_{\rm R}\bigr)m_Q^{}m_q^{}}
{\bigl(\bar s-m_h^2\bigr)^2+\Gamma_h^2m_h^2} ~. ~~~~~~
\end{eqnarray}
For \,$2m_D^{}<m_h^{}<m_Q^{}-m_q^{}$,\, the Higgs-pole contribution dominates this integral, 
and so one has  \,$\Gamma(Q\to qDD)\simeq\Gamma(Q\to qh)\,{\cal B}(h\to DD)$.\,

\begin{figure}[b]
\includegraphics{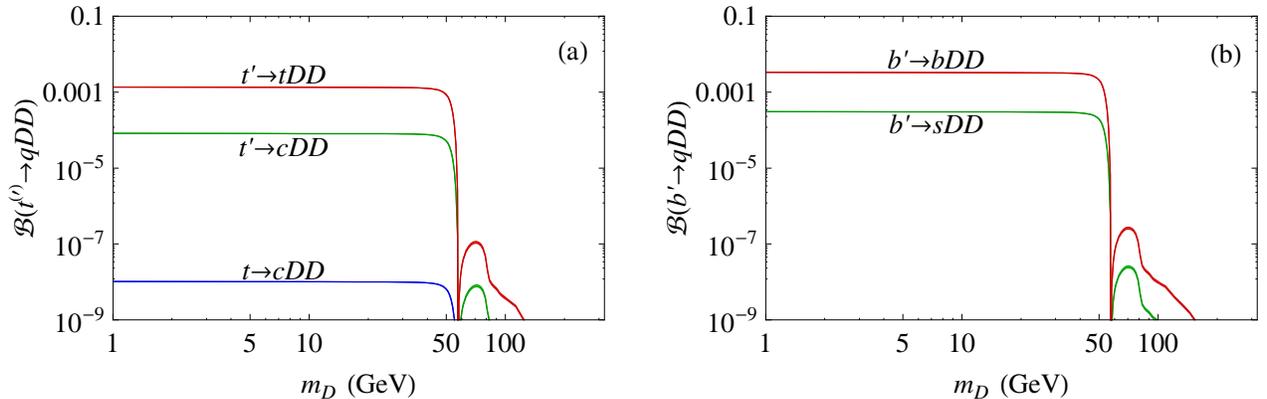}
\caption{Branching ratios of (a)~$t\to cDD$,\, \,$t'\to cDD$,\, and \,$t'\to tDD$\,
and (b)~$b'\to sDD$\, and \,$b'\to bDD$\,
as functions of darkon mass $m_D^{}$ for Higgs mass \,$m_h^{}=115$\,GeV\, in SM4+D
with \,$m_{t'}^{}=500$\,GeV.\label{q2q'dd}}
\end{figure}

The effective couplings $g^{Qq}_{\rm L,R}$ in the SM have been evaluated previously for arbitrary
values of the external and internal masses~\cite{Arhrib:2006pm,Krawczyk:1989qp,Eilam:1989zm}.
We make use of the formulas provided in Ref.~\cite{Eilam:1989zm}.
In our numerical illustration below, we take \,$m_h^{}=115$\,GeV\, and \,$m_{t'}=500$\,GeV,
as well as the corresponding elements of the CKM4 matrix extracted from a global fit in
Ref.~\cite{Eilam:2009hz}.
To determine the branching ratios, we normalize the decay rates according to
\begin{eqnarray}
{\cal B}(t\to cDD) &=& \frac{\Gamma(t\to cDD)}{\Gamma(t\to bW)} ~, \\
{\cal B}(t'\to qDD) &=& \frac{\Gamma(t'\to qDD)}{\Gamma(t'\to bW)+\Gamma(t'\to sW)} ~, \\
{\cal B}(b'\to qDD) &=& \frac{\Gamma(b'\to qDD)}{\Gamma(b'\to tW)+\Gamma(b'\to cW)} ~,
\end{eqnarray}
following Ref.~\cite{Eilam:2009hz} in the \,$Q\to qh$\, cases.
We display the results in Fig.~\ref{q2q'dd}.

Estimates suggest that \,$t\to ch$\, can be detected at the LHC if its branching ratio is
several times $10^{-5}$ or higher~\cite{AguilarSaavedra:2000aj}.
In the presence of the darkon, if \,$h\to DD$\, is leading, \,$t\to c h\to cDD$\, is more
likely to occur than other \,$t\to c h\to c X$\, modes.
In that case, however, as Fig.~\ref{q2q'dd} indicates,
\,${\cal B}(t\to cDD)\;\mbox{\small$\lesssim$}\;1.0\times10^{-8}$\, and so \,$t\to cDD$\,
probably will not be observable in the near future.
The branching ratio could be several times higher if \,$m_{t'}\sim700$\,GeV,\, but this
already exceeds the perturbative unitarity upper-bound \,$m_{t'}\sim550$\,GeV\,~\cite{Holdom:2009rf}.

In contrast, the $t'$ and $b'$ numbers in Fig.~\ref{q2q'dd} are much greater:
\,${\cal B}(t'\to cDD)\;\mbox{\small$\lesssim$}\; 8.2\times10^{-5}$,\,
\,${\cal B}(t'\to tDD)\;\mbox{\small$\lesssim$}\; 1.4\times10^{-3}$,\,
\,${\cal B}(b'\to sDD)\;\mbox{\small$\lesssim$}\; 3.1\times10^{-4}$,\, and
\,${\cal B}(b'\to bDD)\;\mbox{\small$\lesssim$}\; 3.3\times10^{-3}$.\,
Since \,$t'\to qh$\, and \,$b'\to qh$\, decays with branching ratios between $10^{-4}$ and
$10^{-2}$ are expected to be within the reach of the LHC~\cite{Eilam:2009hz,Arhrib:2006pm},
we may expect that these \,$t'\to qDD$\, and \,$b'\to qDD$\, decays would also be detectable
at the LHC despite their final states involving missing energy.
Once they are measured, comparing the results with those from DM direct searches could provide
additional consistency tests for the darkon models.

\section{Conclusions\label{concl}}

We have explored one of the simplest dark-matter models, the SM4+D, consisting of the standard
model with four generations and a real gauge-singlet scalar, the darkon, to play the role of 
WIMP dark matter.
This model possesses not only the phenomenologically interesting features of the SM4,
but also a high degree of predictivity in its DM sector.
We have investigated constraints on the SM4+D from DM direct-search experiments and
from $B$-meson decay into a~kaon plus missing energy.
Compared to the SM3+D case, the resulting bounds in the SM4+D are similar, namely that
for the representative Higgs masses chosen most of the darkon masses between
roughly 4 to 80~GeV are excluded by the direct searches and that much of the mass region
below 1.5\,GeV is also excluded by the $B$ decay data.
Interestingly, the SM4+D as well as the SM3+D can also accommodate the possible interpretation 
that the excess events recently measured by the CDMS\,II and CoGeNT experiments were due to
interactions with a light WIMP of mass around~9\,GeV.
Darkon masses greater than 80\,GeV in the two models are still viable and can be probed 
by future direct searches.

We have discussed the complementarity of DM direct searches and Higgs studies at colliders
in testing the darkon sector of the SM4+D. 
This can be crucial for a relatively light Higgs boson, which may decay substantially 
into the invisible darkons.
Accordingly, we have pointed out that existence of the darkon could lead to the weakening or
evasion of some of the restrictions on the Higgs mass in the presence of fourth-generation
fermions.

We have considered some implications of the SM4+D that are lacking or absent in the SM3+D
as far as probing the darkon properties is concerned.
In particular, we have examined the Higgs-mediated FCNC decays  \,$t\to cDD$,\,
\,$t'\to(c,t)DD$,\, and \,$b'\to(s,b)DD$,\,
which may have observable rates at current or future colliders.
These processes promptly proceed from the \,$Q\to q h$\, transitions if the decay mode
\,$h\to DD$\, is dominant.
Although the \,$t\to cDD$\, branching-ratio is enhanced by several orders of magnitude
compared to that in the SM3+D, reaching the $10^{-8}$ level, this decay is still unlikely 
to be measurable in the near future.
In~contrast, the branching ratios of \,$t'\to qDD$\, and \,$b'\to qDD$\, can be as large as
a few times $10^{-3}$, which may be detectable at the LHC.
If observed, they would offer extra means to test the models, covering darkon masses from zero
up to hundreds of~GeV.

\acknowledgments \vspace*{-1ex}
This work was partially supported by NSC and NCTS.

\end{document}